\documentclass[aps,prl,twocolumn,showpacs,superscriptaddress]{revtex4-1}

\usepackage{graphicx}
\usepackage{amsmath}
\usepackage{bm}

\newcommand{\BFO}{BiFeO$_3$}
\begin{document}

\title{X-ray imaging and multiferroic coupling of cycloidal magnetic domains in ferroelectric monodomain BiFeO$_3$}
\author{R. D. Johnson}
\affiliation{Clarendon Laboratory, Department of Physics, University of Oxford, Oxford, OX1 3PU, United Kingdom}
\affiliation{ISIS facility, Rutherford Appleton Laboratory-STFC, Chilton, Didcot, OX11 0QX, United Kingdom}
\author{P. Barone}
\affiliation{Consiglio Nazionale delle Ricerche, Istituto Superconduttori, materiali innovativi e dispositivi (CNR-SPIN),
67010 L'Aquila, Italy}
\author{A. Bombardi}
\affiliation{Diamond Light Source, Harwell Science and Innovation Campus, Didcot, OX11 0DE, United Kingdom}
\author{R. J. Bean}
\affiliation{C.M.M.P., Department of Physics and Astronomy, University College London, Gower Street, London WC1E 6BT, United Kingdom}
\author{S. Picozzi}
\affiliation{Consiglio Nazionale delle Ricerche, Istituto Superconduttori, materiali innovativi e dispositivi (CNR-SPIN),
67010 L'Aquila, Italy}
\author{P. G. Radaelli}
\affiliation{Clarendon Laboratory, Department of Physics, University of Oxford, Oxford, OX1 3PU, United Kingdom}
\author{Y. S. Oh}
\author{S-W. Cheong}
\affiliation{Rutgers Center for Emergent Materials and Department of Physics and Astronomy, 136 Frelinghuysen Road, Piscataway 08854, New Jersey, USA.}
\author{L. C. Chapon}
\affiliation{Institut Laue-Langevin, BP 156, 6, rue Jules Horowitz, 38042 Grenoble Cedex 9, France}

\begin{abstract}
Magnetic domains at the surface of a ferroelectric monodomain \BFO\ single crystal have been imaged by hard X-ray
magnetic scattering. Magnetic domains up to several hundred microns in size have been observed, corresponding to
cycloidal modulations of the magnetization along the wave-vector $\boldsymbol{k}$=($\delta$,$\delta$,0) and
symmetry equivalent directions. The rotation direction of the magnetization in all magnetic domains, determined by diffraction of circularly polarized light, was found to be unique and in agreement with 
predictions of a combined approach based on a spin-model complemented by relativistic density-functional simulations. Imaging of the surface shows that the largest adjacent domains display a 120$^{\circ}$ vortex structure.
\end{abstract}


\maketitle

\begin{figure}
\includegraphics[width=\columnwidth]{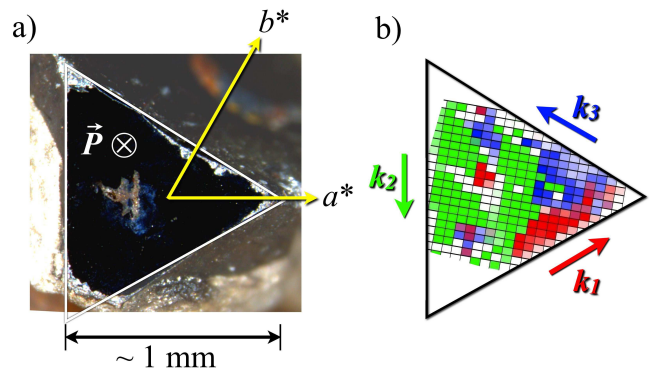}
\caption{\label{fig:2} (Color online) a) Photograph of the polished crystal surface of \BFO\ normal to the (001) axis (hexagonal setting, see text for details). A v-shaped defect is seen in the center of the surface. The downward direction of the electric polarization $\boldsymbol{P}$ determined by PFM is shown (cross) together with the reciprocal $a^*,b^*$ axis (yellow lines). b) Distribution of antiferromagnetic domains with wave-vectors $\boldsymbol{k_1}=(\delta,\delta,0)$ (red), $\boldsymbol{k_2}=(\delta,-2\delta,0)$ (green) and $\boldsymbol{k_3}=(2\delta,-\delta,0)$ (blue). The direction of propagation of the cycloidal modulation in real-space coordinates for each domain is shown. Each pixel is colored according to the diffraction signal (domain) present. In the case of multiple diffraction peaks (overlap of domains), the pixels are shaded with mixed colors, respectively.}
\end{figure}

\begin{figure}
\includegraphics[width=\columnwidth]{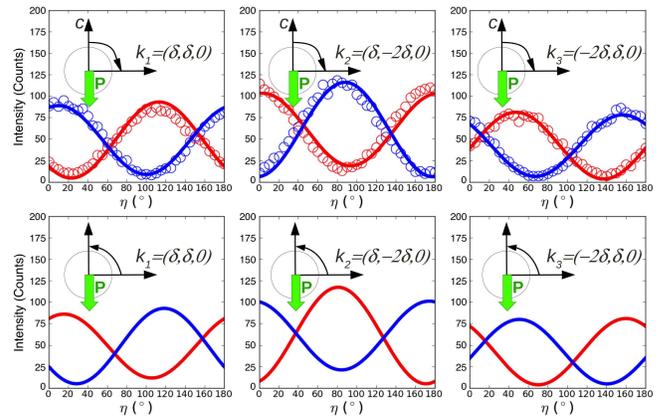}
\caption{\label{fig:3} (Color online) Top: Variation of the scattered X-ray intensity with the analyser angle $\eta$
(circle symbols) for three magnetic reflections ($\delta$,$\delta$,0), ($\delta$,-2$\delta$,0), (-2$\delta$,$\delta$,0). The red (blue) color corresponds to the signal observed with a right-handed (left-handed) X-ray incident polarization. The solid lines show the results of a least-square refinement of the BiFeO$_3$ magnetic structure assuming $\beta$=-1 (CW, see text for details). Bottom: Calculated variation of the scattered X-ray intensity with the analyser angle $\eta$  assuming $\beta$=+1 (CCW, see text for details). The direction of electric polarization $\boldsymbol{P}$ is shown as a green arrow. }
\end{figure}

The seminal work of I. Dzyaloshinsky \cite{I1958241} on the relativistic origin of weak ferromagnetism in antiferromagnetic substances is intimately connected to various emergent physical phenomena in condensed matter. For example, in the skyrmion lattice the very presence of antisymmetric exchange interactions (Dzyalonshinskii-Moriya \cite{I1958241,PhysRev.120.91}) in a non-centrosymmetric crystal stabilizes the long period helical structure in zero magnetic field. Also, for some spin-driven ferroelectrics (multiferroics), the electric polarization is driven by non-collinear magnetic orders; the \emph{inverse} Dzyalonshinskii-Moriya effect. In this case, a phenomenological formulation \cite{PhysRevLett.96.067601} shows that for cycloidal magnetic structures, i.e. spins rotating in a plane that contains the magnetic wave-vector ($\boldsymbol{k}$), the electric polarization ($\boldsymbol{P}$) transforms as a product involving the magnetization density and its gradient; the so-called Lifshitz invariant of the form $\boldsymbol{P} \cdot \boldsymbol{\lambda}$, where $\boldsymbol{\lambda}=(\boldsymbol{\nabla} \cdot \boldsymbol{L})\boldsymbol{L}-(\boldsymbol{L} \cdot \boldsymbol{\nabla}) \boldsymbol{L}$, and $\boldsymbol{L}$ is the antiferromagnetic order-parameter. In a complementary view, the magnetic polarity can be thought of as arising locally from spin current \cite{knb}, as $\boldsymbol{\lambda}=\boldsymbol{k}\times(\boldsymbol{S}_i\times\boldsymbol{S}_j)$, where $\boldsymbol{S}_i$ and $\boldsymbol{S}_j$ are spins on adjacent sites.  Like $\boldsymbol{P}$, $\boldsymbol{\lambda}$ is a polar vector, and will be called \emph{magnetic polarity} in the remainder.  \\
\indent In \BFO, arguably the most studied multiferroic owing to room temperature magnetoelectric coupling \cite{Zhao2006}, ferroelectricity is the consequence of an improper structural transition at T$_c$ $\sim$ 1100K to the polar space group $R3c$. In bulk samples, the magnetic ordering transition occurs at T$_N \sim$ 640K. While the two do not coincide, the respective order parameters are coupled through antisymmetric exchange, i.e.,  $\boldsymbol{P}$ drives the appearance of the inhomogeneous magnetization through a coupling term $\gamma \boldsymbol{\lambda}\boldsymbol{P}$, where $\gamma$ is a coupling constant, a scenario originally proposed by Kadomtseva \cite{springerlink:10.1134/1.1787107}. The magnetic structure can be described locally as canted G-type, but with a long period modulation ($\sim$ 620 $\mathrm{\AA}$) in the hexagonal basal plane \cite{0022-3719-15-23-020}. Subsequent studies \cite{PhysRevLett.100.227602,PhysRevB.78.100101} determined that the modulation is cycloidal with the spins rotating in the $(\boldsymbol{k},\boldsymbol{z})$-plane where $\boldsymbol{k}$ can take the three symmetry-equivalent directions $\boldsymbol{k_1}=(\delta,\delta,0)$, $\boldsymbol{k_2}=(\delta,-2\delta,0)$ and $\boldsymbol{k_3}=(-2\delta,\delta,0)$ in the hexagonal setting of the $R3c$ group (employed throughout), and $\delta$=0.0045 at 300K. \\
\indent In this letter, we study the magnetic domains at the surface of a millimeter-size single crystal of \BFO\ with a single ferroelectric (FE)  domain. Using the high momentum and spatial resolution of synchrotron X-ray diffraction, combined with circular polarization of the beam, we determine the absolute rotation direction of the magnetization in individual magnetic domains, which are found to have the same magnetic polarity. The sign of $\gamma$ is determined and compared to model-Hamiltonian and $ab-initio$ calculations. The large domains observed appear to form vortex structures with a closure of the wave-vector for three adjacent 120$^{\circ}$ domains. \\
\indent Single crystals of several mm$^3$ were grown from a Bi$_2$O$_3$/Fe$_2$O$_3$/B$_2$O$_3$ flux by slow cooling from 870 $^{\circ}$C to 620 $^{\circ}$C. A selected crystal was mechanically cut and polished perpendicular to the $c$-axis, and then annealed to remove any induced strain. A piezoresponse force microscopy (PFM) of the polished face (not shown) indicated that the surface had a single FE domain, with the electrical polarization pointing down into the sample. We label this domain FE$\downarrow$, with the opposite polar domain labelled FE$\uparrow$. The synchrotron X-ray experiments were performed at Diamond Light Source (UK) on Beamline I16 \cite{I16}. A horizontally polarized beam with a flux of $\sim10^{12}$ photons per second was delivered by a linear undulator and tuned to an energy of 5.8 keV, off resonance of chemical elements present in \BFO. Circular polarization of the beam was achieved by transmission through a 100$\mu m$ thick diamond phase-plate, reducing the incident flux by $\sim40~\%$. The diamond crystal was aligned to scatter near the (111) reflection in transmission. For a certain deviation of $\Delta\theta$ from the Bragg condition, the crystal behaves as a quarter wave plate giving circular light. The handedness of the light is determined by the sign of $\Delta\theta$, which was calculated by dynamical scattering theory, and confirmed through experimental calibration of the beam line by measuring the X-ray dichroism of a standard ferromagnet.\\
\begin{figure}
\includegraphics[width=\columnwidth]{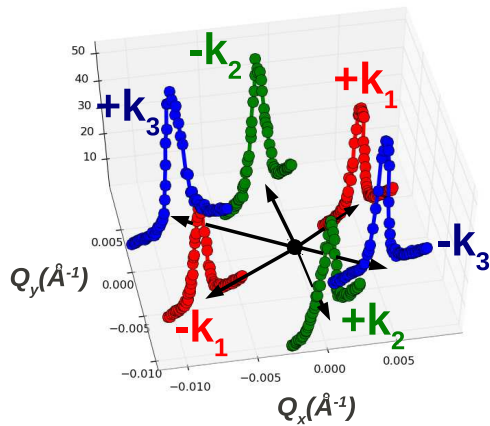}
\caption{\label{fig:1} (Color online) Reciprocal-space scans showing magnetic Bragg intensities of the six satellites of the
(0,0,9) parent reflection, with $\boldsymbol{k_1}=(\delta,\delta,0)$ (red), $\boldsymbol{k_2}=(\delta,-2\delta,0)$
(green) and $\boldsymbol{k_3}=(-2\delta,\delta,0)$ (blue) and $\delta \sim$ 0.0045. The $x$- and $y$-axis are taken respectively along the reciprocal $a^*$ direction and real space $b$-direction.}
\end{figure}
\indent To prevent contamination of the magnetic signal by charge scattering from neighboring structural reflections ($\delta$ is extremely small), we focussed on magnetic satellites of the $\boldsymbol{N}$=(0,0,9) reflection, which is extinct by the presence of \textit{c}-glide planes. Additionally, contamination from multiple scattering was fully eliminated by positioning the sample at an azimuthal angle $\phi$=-170.0$^{\circ}$ with respect to [1,0,0]. Diffraction of $\lambda /2$ X-rays was made negligible by employing up-stream harmonic rejection mirrors. The magnetic signal was clearly identified using the full X-ray beam size (100$\mu m$ vertical x 350$\mu m$ horizontal) with linearly-polarized light scanning in reciprocal-space around the positions of the six satellites $\boldsymbol{N}\pm\boldsymbol{k_1}$,$\boldsymbol{N}\pm\boldsymbol{k_2}$,$\boldsymbol{N}\pm\boldsymbol{k_3}$, for various positions on the crystal surface. The high momentum resolution allows the full separation of the six satellites, shown in Fig. \ref{fig:1}, in contrast to previous neutron experiments \cite{PhysRevLett.100.227602,PhysRevB.78.100101}.
The beam size was subsequently reduced using slits to create a footprint of 50x50$\mu m^2$ on the crystal surface. An image of the magnetic domains (Fig. \ref{fig:2}) was then constructed by step-scanning the sample position with a step size of $50\pm$ 1$\mu$m, recording the intensities of magnetic Bragg peaks $\boldsymbol{N}+\boldsymbol{k_1}$,$\boldsymbol{N}+\boldsymbol{k_2}$,$\boldsymbol{N}+\boldsymbol{k_3}$ using rocking-curve scans. This procedure lead to the identification of three large magnetic $k$-domains corresponding to $\boldsymbol{k_1}$, $\boldsymbol{k_2}$ and $\boldsymbol{k_3}$, shown in Fig. \ref{fig:2}, and to some smaller domains at the edges of the scanned surface and around a sizeable crystal imperfection in the center of the specimen. The three main domains are extremely large, reaching up to 500 $\mu m$ in some directions. Note that the average penetration depth of the X-ray beam is 3.3 $\mu$m at this energy, placing a lower bound on the domain thickness. Despite the long period of the modulation (620 $\mathrm{\AA}$), this result indicates that each domain corresponds to several hundred magnetic periods. The real space directions of the wave-vectors are shown in Fig. \ref{fig:2}b. It appears that the modulation of the magnetization follows a 120$^{\circ}$ vortex structure described by the path $\boldsymbol{k_1}\rightarrow\boldsymbol{k_3}\rightarrow\boldsymbol{k_2}$ when rotating anticlockwise on the crystal surface.\\     
\indent To determine the absolute rotation direction of the magnetization in each domain (magnetic polarity), scattering data were collected using circularly polarized light. For alternate chiralities of the X-ray beam (left/right handed), the intensities of the magnetic signals $\boldsymbol{N}+\boldsymbol{k_1}$, $\boldsymbol{N}+\boldsymbol{k_2}$, and $\boldsymbol{N}+\boldsymbol{k_3}$, were recorded after analysis with a pyrolitic graphite crystal as a function of the analyzer angle $\eta$, where $\eta$=0 and $\eta$=90 correspond to the $\sigma'$ and $\pi'$ polarization channels (perpendicular and parallel to the scattering plane), respectively. The incident-light polarization is described by the Stokes vector $\boldsymbol{P_s}$=($P_1$,$P_2$,$P_3$) \cite{deBergevin:a19658}, where $P_1$, $P_2$, $P_3$ represent respectively the degree of linear polarization along $\sigma$ and $\pi$, oblique polarization ($\pm45^{\circ}$) and left or right circular polarization. $P_1$ and $P_2$ have been determined by fitting the variation with $\eta$ of the Thomson scattering intensity for the reflection (0,0,6), taking into account the cross-channel leakage of the analyzer. $|P_3|$ was determined by supposing a fully polarized beam, i.e. $|P_3|=\sqrt{1-P_1^2-P_2^2}$. In our measurements, right and left handed light was 93\%\ and 92\%\ circularly polarized, respectively (see supplementary information for the detailed calculations and conventions used). For each magnetic domain, the intensity ($I_M$) of the corresponding diffraction peak was evaluated using the density-matrix formalism \cite{Fano1957}:
\begin{equation}
I_M(\boldsymbol{Q},\boldsymbol{P_s},\eta)=tr[D(\eta).V_m(\boldsymbol{Q}).\rho(\boldsymbol{P_s}).V_m(\boldsymbol{Q})^\dagger]
\end{equation}  
where $\rho$ is the density-matrix representing the polarization of the incident beam, and $D$ the matrix representing the analyser configuration. $V_m$=$\boldsymbol{B.M(Q)}$ is the scattering amplitude where $\boldsymbol{B}$ is expressed as a two by two matrix on the basis of the $\sigma$ and $\pi$ polarizations \cite{deBergevin:a19658} and $\boldsymbol{M(Q)}$ the magnetic unit-cell structure factor. For the peaks at $\boldsymbol{Q}=(0,0,9)+\boldsymbol{k_i}$ ($i$=1,2,3): 
\begin{equation}
\boldsymbol{M(Q)}=6f(\boldsymbol{Q})[\boldsymbol{M_{\|}}-\beta i \boldsymbol{M_{z}}].e^{-i.18 \pi z}
\end{equation} 
where f($\boldsymbol{Q}$) is the magnetic form factor for Fe$^{3+}$, calculated in the dipolar approximation from \cite{MagneticFormFactors}, $\boldsymbol{M_{\|}}$ and $\boldsymbol{M_{z}}$ are the magnetization vectors of the cycloid along $\boldsymbol{k_i}$ and the $c$-axis, respectively, and $z$ is the fractional coordinate of Fe in the unit-cell ($z$=0.2208  at 300K). In our conventions $\beta=+1$ and $\beta$=-1 correspond to cycloids rotating counterclockwise (CCW) and clockwise (CW), respectively, when the structure is viewed propagating along $\boldsymbol{k_i}$ and $c$ is up.\\ 
\indent Comparison of intensities collected on the three main domains and calculations assuming circular cycloids (Fig. \ref{fig:3}), unambiguously demonstrates that all magnetic configurations rotate CW following our definition. This is inferred from the $\eta$-positions of the $I_m$ extrema obtained with both light polarizations, which would be interchanged for a structure of opposite magnetic polarity. Within our conventions, $\boldsymbol{\lambda}$ is oriented in the $+c$ direction, antiparallel to $\boldsymbol{P}$.  Refining the ellipticity of the cycloid ($\boldsymbol{M_{z}}/\boldsymbol{M_{\|}}$) does not lead to significant improvements. This, and the failure to observe higher order magnetic satellites, supports the picture of a harmonic modulation at 300K, discussed in \cite{PhysRevB.84.144404,PhysRevB.83.174434}. No improvements of the fit were obtained by considering a slight tilt of the cycloidal plane, as recently suggested \cite{PhysRevLett.107.207206}. \\    

\begin{figure}
\includegraphics[width=\columnwidth]{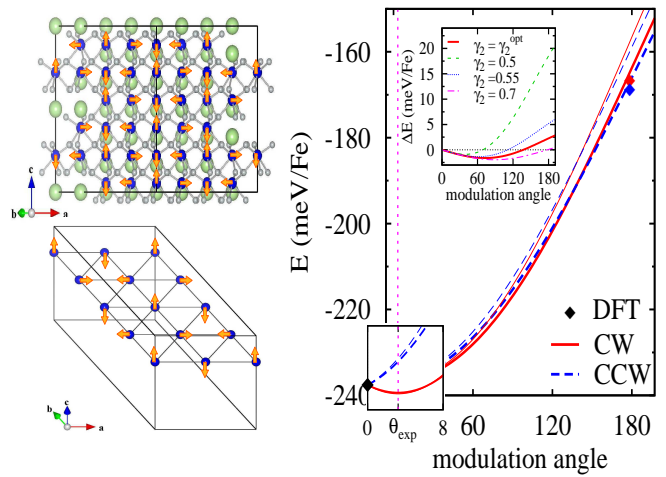}
\caption{(Color online) Sketch of the considered magnetic configuration in the $2a$~x~$4b$~x~$c$ hexagonal cell of  BiFeO$_3$. Upper left: side view. Bottom left:  spin configuration for  a selected layer of Fe ions. Right panel: mean-field energy as a function of the modulation angle for the FE$\downarrow$ domain, with all the parameters estimated from DFT  with $U$=5 eV (see text); vertical dotted line marks the experimental
$\theta_{exp}$, thick (thin) lines corresponds to total energy with (without) next-nearest neighbor contribution $J_{nnn}$. A zoom for small modulation angles is also shown. The inset shows the energy difference between CW and CCW configurations for the optimal $\gamma,\gamma_2$ and by artificially modifying the $nnn$ contribution $\gamma_2$.}
\label{fig:dft}
\end{figure}

\indent  The relationship between ferroelectric and magnetic polarity was further investigated through $ab-initio$ spin-constrained calculations in the framework of density-functional theory (DFT). The VASP code \cite{vasp} with the PAW pseudopotentials \cite{paw} was employed within the GGA+$U$ approach \cite{pbe,ldau} ($U$ ranging between 3 and 7 eV and $J$=1 eV for Fe $d$-states) including spin-orbit coupling, with a plane-wave cutoff of 450 eV. The total polarization was calculated via the Berry-phase formalism \cite{vanderbilt,resta}. Structural parameters for the FE phase were taken from Ref. \cite{palewicz}. Due to its long periodicity, the true modulation of the magnetization is currently unaccessible to DFT. The modulation angle of the antiferromagnetic order parameter is given by $\theta=2\pi(q_x x+q_y y)$, where ${\bm q} = {\bm k}_1,{\bm k}_2$, or ${\bm k}_3$. Choosing ${\bm q} ={\bm k}_3$, corresponding to a cycloidal modulation of spins rotating in the $ac$-plane, one needs a supercell $na$~x~$2nb$~x~$c$ in order to accommodate $\theta = 2\pi/na$. The largest possible supercell, $2a$~x~$4b$~x~$c$, contains 240 atoms (just within the capabilities of state-of-the-art DFT simulations) and has modulation angle $\pi/a$. Accordingly, we considered a hypothetical spin configuration where the cycloidal period is reduced to two unit cells along $a$, with spins rotating CW (see Fig. \ref{fig:dft}, left panels) or CCW. The total energies of the two states are then compared in two symmetry-equivalent FE states with opposite polarization, and in a reference paraelectric (centrosymmetric, $R\bar{3}c$) structure.
\begin{table}
\caption{DFT results obtained for $U$=5 eV, $J$=1 eV. The energy difference is defined as $\Delta E = E_{CW}-E_{CCW}$.
FE$\uparrow$ and FE$\downarrow$ are characterized by opposite collective displacements, $\tau$, respectively upward and downward, of Bi sublattice
with respect to O layers perpendicular to $c$ axis.}\label{tab1}
\begin{ruledtabular}
\begin{tabular}{c|cccc}
& $\tau$ (\AA)& $P_c$\,($\mu$C/cm$^2$)          &   $\Delta$ E\,(meV/Fe) & Favored rotation\\ 
\hline
FE$\uparrow$ &  0.668    & 105.17        &  -2.34  & CW\\
PE  &  0                &  0            &  0     & -  \\
FE$\downarrow$ &  -0.668   &  -105.17      &  2.34 & CCW\\
\end{tabular}
\end{ruledtabular}
\end{table}
As shown in Table \ref{tab1}, the paraelectric state is degenerate with respect to magnetic polarity, which is then lifted in both FE states. Furthermore, the energy favored state switches when polarization is switched. The reliability of this trend has been checked for different values of U, as well as within a conventional local-density approximation, giving $\vert\Delta E\vert$ between 1.1 and 4.7 meV/Fe. These findings strongly point to a tight relationship between the magnetic polarity of the cycloidal modulation and the FE polarization. However, the rather large energy difference $\Delta E$, as well as the
disagreement of the predicted magnetic polarity with the experimental finding, are most probably due to the artificially short modulation of the magnetic configuration imposed in DFT calculations. Testing this hypothesis by mapping the energy evolution as a function of the modulation vector would require very demanding - if at all possible - DFT calculations.  Instead, we adopted a different strategy as follows.

We introduce a Heisenberg-like spin model with nearest neighbor (\emph{nn}) and next nearest neighbor (\emph{nnn}) symmetric, as well as antisymmetric exchange interactions. The symmetric exchange interactions have been estimated by mapping the DFT energy of collinear ferro- and antiferromagnetic spin configurations onto the Heisenberg model, giving $J_{nnn}/J_{nn}\sim 0.03$, consistent with the value extrapolated from spin-wave dispersions \cite{matsuda,jeong}. The antisymmetric exchange interactions for a given direction of $\boldsymbol{P}$ are captured through the magnetoelectric coupling constants $\gamma$ (\emph{nn}) and $\gamma_2$ (\emph{nnn} weight). $\gamma$ and $\gamma_2$ are then estimated by imposing the following constraints on the mean-field Heisenberg energy: i) the minimum of the energy occurs at the experimental modulation angle $\theta_{exp}\sim 3.24^\circ$ and ii) the energy difference at $\theta=\pi$ (i.e. the spin configuration simulated in our DFT calculations) is equal to $\Delta E$, as evaluated from first principles. Under these assumptions we can estimate $\gamma\simeq 2.38 \cdot 10^{-4} V$ and
$\gamma_2^{opt}=0.6$, with $\vert\Delta E(\theta_{exp})\vert\simeq 0.11~ meV$/Fe, for $U=5$ eV (the same order of magnitude was obtained for $U=3$ eV and $U=7$ eV). Following Ref. \cite{springerlink:10.1134/1.1787107}, the inhomogeneous magnetoelectric
coefficient in the  framework of Landau theory of phase transitions would be
$\gamma=4\pi A/l P_c\sim 5.8 \cdot 10^{-4} V $ (with exchange stiffness $A=1.87\cdot 10^{5}$ eV/cm \cite{springerlink:10.1134/1.1787107}, modulation period $l=620$~\AA\, and assuming the calculated $P_c$ = 105.17\,$\mu$C/cm$^2$), in good qualitative agreement with our estimate. Our model analysis also underlines the relevant role of $nnn$ interactions, as through including $J_{nnn}$ the mean-field Heisenberg energy almost reproduces the DFT results even at $\theta=\pi$, where the only constraint has been imposed on $\Delta E$ (Fig. \ref{fig:dft}). As anticipated, the energy-favored magnetic polarity appears to depend strongly on the modulation angle of the cycloidal configuration and on the relative weight of $nn$ and $nnn$ antisymmetric exchange interactions, which give rise to opposite energy contributions with a different dependence on $\theta$ (as detailed in the supplementary information). For $\gamma\lesssim 0.7$, the energetic competition between $nn$ and $nnn$ interactions causes the favored magnetic polarity to change sign when moving from short to long modulation periods, therefore reconciling DFT and experimental results.

\indent In summary, magnetic domains of up to 500 $\mu$m have been observed at the surface of a single crystal of BiFeO$_3$ consisting of a single ferroelectric domain. The magnetic cycloids in each domain were found to propagate with a unique rotation direction imposed by the electric polarity of the crystal, in agreement with the predictions of our theoretical study if \emph{nnn} interactions are taken into account. In future studies, it would be of interest to observe the switching of the rotation direction of the magnetic cycloids upon switching of the ferroelectric polarization by an applied electric field, as observed in TbMnO$_3$ \cite{fabrizi09}, and predicted by our calculations.

\begin{acknowledgements}
The work done at the University of Oxford was funded by an EPSRC grant, number EP/J003557/1, entitled ``New Concepts in Multiferroics and Magnetoelectrics", and the work at Rutgers was supported by DOE DE-FG02-07ER46328. Work in L'Aquila was supported by the European Research Council  (ERC-StG No.203523 BISMUTH) and by the CARIPLO Foundation (No. 2010-0584 ECOMAG).
\end{acknowledgements}

\bibliography{BiFeO3_ps}

\end{document}